\documentclass{elsart-alt} 
\usepackage{amssymb,amsfonts,amsmath,latexsym,graphicx}
\newcommand{\bq}{\begin{equation}}
\newcommand{\eq}{\end{equation}}
\newcommand{\eqn}[1]{(\ref{#1})}
\newcommand{\Sum}{\displaystyle\sum}

\newcommand{\Int}{\displaystyle\int}
\newcommand{\Frac}{\displaystyle\frac}

\newcommand{\Sup}{\displaystyle\sup}
\newcommand{\Lim}{\displaystyle\lim}

\newcommand{\R}{\mathbb{R}}
\newcommand{\C}{\mathbb{C}}
\newcommand{\cal}{\mathcal}
\newcommand{\proof}{\noindent{\sl Proof.\/} }
\newcommand{\finprf}{\unskip\null\hfill$\square$\vskip 0.3cm}
\newcommand{\Id}{{\rm Id}}
\newcommand{\ds}{\displaystyle}

\newtheorem{theorem}{Theorem}
\newtheorem{proposition}[theorem]{Proposition}
\newtheorem{lemma}[theorem]{Lemma}
\newtheorem{corollary}[theorem]{Corollary}
\newenvironment{remark}{\par\medskip\noindent{\bf Remark }\it }{\/\par}
\begin{document}
\begin{frontmatter}
\title{An analytical proof of Hardy-like inequalities related to the Dirac operator\thanksref{copyright}}

\thanks[copyright]{\copyright\, 2003 by the authors. This paper may be reproduced, in its entirety, for non-commercial purposes.}

\author[Dauphine]{Jean Dolbeault\thanksref{Europe}}
\author[Dauphine]{Maria J. Esteban\thanksref{Europe}}
\author[Atlanta]{Michael Loss\thanksref{USA}}
\author[Bilbao]{Luis Vega\thanksref{Europe}}

\address[Dauphine]{CEREMADE, Universit\'e Paris-Dauphine, 75775 Paris Cedex 16, France}
\address[Atlanta]{School of Mathematics, Georgia Tech, Atlanta, GA 30332, USA}
\address[Bilbao]{Universidad del Pa\'{\i}s Vasco, Departamento de Matem\'aticas, Facultad de Ciencias, Apartado 644, 48080 Bilbao, Spain}

\thanks[Europe]{PICS (CNRS, Paris-Bilbao) and European Programs HPRN-CT \# 2002-00277 \& 00282.}
\thanks[USA]{Work partially supported by U.S. National Science Foundation grant DMS 00-70589.}

\begin{abstract} We prove some sharp Hardy type inequalities related to the Dirac operator by elementary, direct methods. Some of these inequalities have been obtained previously using spectral information about the Dirac-Coulomb operator. Our results are stated under optimal conditions on the asymptotics of the potentials near zero and near infinity. \end{abstract}

\end{frontmatter}

\section{Introduction and main results}\label{S1}

The {\sl uncertainty principle\/} is without any doubt a fundamental attribute of quantum mechanics \cite{Reed-Simon78}. In the case of the Laplacian it states that for all functions $f \in C^{\infty}_0(\R^3)$,
\begin{equation}\label{uncertainty}
\int_{\R^3}|{\boldsymbol{\nabla}}f|^2\,dx\geq\frac{1}{4}\int_{\R^3}\frac{|f|^2}{|x|^2}\,dx\;.
\end{equation}
This inequality is also known as Hardy's inequality. By scaling, the power of the potential is seen to be optimal but also the constant $\frac{1}{4}$ cannot be improved. However, it is still possible to improve the inequality by adding lower order terms.

In recent years there has been a great effort to find optimal improved Hardy-type inequalities in the case of the Laplacian. The pioneering work in this direction is due to Brezis and V\'azquez \cite{Brezis-Vazquez-96} in the case of Dirichlet boundary conditions, and to Lieb and Yau \cite{Lieb-Yau-88} in the case without boundary conditions. Further improvements have been obtained in \cite{Adimurthi,Adimurthi-Esteban-03,Adimurthi-Ramaswamy-Chaudhuri-01,Adimurthi-Sandeep-01,Barbatis-Filippas-Tertikas-1,Barbatis-Filippas-Tertikas-2,Filippas-Tertikas-02,Sandeep-01}. 

An analogue of this inequality for a relativistic version of the Schr\"odinger equation where the Laplacian is replaced by $\sqrt{-\Delta}$ is an inequality due to~Kato~:
\begin{equation}\label{Kato}
\left(f,\sqrt{-\Delta}f\right)\geq\frac{2}{\pi}\int_{\R^3}\frac{|f|^2}{|x|}\,dx
\end{equation}
(see \cite{Herbst-77,Kato-66}). In this inequality the power and constant are again optimal. An immediate consequence of this inequality is that the relativistic model of the hydrogenic atom with kinetic energy $\sqrt{-\Delta}$ and with nuclear charge $Z$ is {\sl stable\/} if and only if $\nu:= Z\alpha \leq \pi/2$, where $\alpha\approx 1/137$\dots is the fine structure constant. Lieb and Yau \cite{Lieb-Yau-88} also discovered some generalizations for $\sqrt{-\Delta}$ to balls. 

The Dirac relativistic hydrogenic atom is {\sl stable\/} only if $\nu <1$. The Dirac Hamiltonian is unbounded from below and {\sl instability\/} has to be interpreted in the different, more subtle, sense of a breakdown of selfadjointness of the Dirac-Coulomb Hamiltonian. If the Coulomb singularity is smeared out, then the threshold for stability is reached when the lowest eigenvalue in the gap reaches the upper bound of the negative continuum. This happens in general for larger values of $\nu$. In the case of the Dirac-Coulomb Hamiltonian, the stability is a consequence of the following Hardy-type inequality.
\begin{theorem}\label{TR1} {\rm\cite{Dolbeault-Esteban-Sere-00B}} Let $\boldsymbol{\sigma}=(\sigma_i)_{i=1,2,3}$ be the Pauli-matrices :
\begin{center}
$\sigma_1={\left(\begin{smallmatrix}0 \;&\; 1 \\ 1 \;&\; 0 \end{smallmatrix}\right)}\;$, \quad
$\sigma_2={\left(\begin{smallmatrix}0 \;&\; -i \\ i \;&\; 0 \end{smallmatrix}\right)}\;$, \quad
$\sigma_3={\left(\begin{smallmatrix}1 \;&\; 0 \\ 0 \;&\; -1 \end{smallmatrix}\right)}\;$. 
\end{center}
Then for every $\varphi \in H^1 (\R^3, \C^2)$,
\bq\label{R1}
\Int_{\R^3}\left(\Frac{|{\boldsymbol{\sigma}}\cdot{\boldsymbol{\nabla}}\varphi|^2}{1+\frac{1}{|x|}}+|\varphi|^2\right)\,dx \ \geq \ \Int_{\R^3}\Frac{|\varphi|^2}{|x|}\,dx\;.
\eq
\end{theorem}
As in \eqn{uncertainty} and \eqn{Kato}, the powers of $|x|$ and the constants are optimal. Inequality~(\ref{R1}) has been established using a characterization of the eigenvalues of a self-adjoint operator in a gap of its essential spectrum by means of a particular min-max. See below and refer to \cite{Dolbeault-Esteban-Sere-00B} for more details. For other results on min-max characterizations of the eigenvalues of Dirac operators, see \cite{Esteban-Sere-97,Griesemer-Siedentop-99,Griesemer-Lewis-Siedentop-99,Dolbeault-Esteban-Sere-00A}. 

By scaling, if we replace $\varphi(\cdot)$ by $\varepsilon^{-1}\varphi(\varepsilon^{-1}\cdot)$ and take the limit $\varepsilon\to 0$, (\ref{R1}) implies that for all $\varphi\in H^1(\R^3,\C^2)$,
\bq\label{R2}
\Int_{\R^3}|x|\ |{\boldsymbol{\sigma}}\cdot{\boldsymbol{\nabla}}\varphi|^2\,dx\ 
\geq\ \Int_{\R^3}\Frac{|\varphi|^2}{|x|}\,dx\;.
\eq
This inequality is slightly generalized form (take $\varphi=(g,0)$ and consider independently the cases where $g$ takes either real or purely imaginary values) of the following inequality : for all $g\in H^1(\R^3,\C)$, 
\bq\label{R3}
\Int_{\R^3}|x|\ |{\boldsymbol{\nabla}}g|^2\,dx\geq\Int_{\R^3}\Frac{|g|^2}{|x|}\,dx\;, 
\eq
which is itself equivalent to (\ref{uncertainty}) : take $f=\sqrt{|x|}\,g$. Note that the largest space in which Inequality (\ref{R1}) holds is larger than $H^1(\R^3,\C^2)$ and contained in $H^{1/2}(\R^3,\C^2)$. For more details, see \cite{Dolbeault-Esteban-Sere-00B}.

In \cite{Dolbeault-Esteban-Sere-00B}, the proof of (\ref{R1}) has been carried out by using explicit knowledge on the point-spectrum of the Dirac-Coulomb operator $H_\nu:=-i\ {\boldsymbol{\alpha}}\cdot{\boldsymbol{\nabla}}+\beta-\frac{\nu}{|x|}$, where the matrices $\beta$, $\alpha_k \in{\cal M}_{4\times 4}(\C)$, $k= 1$, $2$, $3$, are defined by
\begin{center}
$\alpha_k={\left(\begin{smallmatrix}0 \;&\; \sigma_k \\ \sigma_k \;&\; 0 \end{smallmatrix}\right)}\;$, \quad
$\beta={\left(\begin{smallmatrix}\Id \;&\; 0 \\ 0 \;&\; -\Id \end{smallmatrix}\right)}\;$.
\end{center}
$\Id$ is the identity matrix in $\C^2$ and $\nu$ is a real parameter taking its values in $(0,1)$. It is well-known \cite{Thaller-92} that for any $\nu\in(0,1)$ $H_\nu$ can be defined as a self-adjoint operator with domain ${\cal D}_\nu$ satisfying : $H^1(\R^3,\C^4)\subset{\cal D}_\nu\subset H^{1/2}(\R^3,\C^4)$ and spectrum 
$$
\sigma (H_\nu)\ =\ \sigma_{ess}(H_0)\cup\Big\{\lambda_1^\nu,\,\lambda_2^\nu,\cdots\Big\}\,,\quad\sigma_{ess}(H_0)\ =\ (-\infty,-1]\cup[1,+\infty)\;, 
$$
where $\{\lambda_k^\nu\}_{k\geq 1}$ is the nondecreasing sequence of eigenvalues of $H_\nu$, all contained in the interval $(0,1)$ and such that :
$$
\lambda_1^\nu=\sqrt{1-\nu^2}\,,\quad\Lim_{k\rightarrow +\infty}\lambda_k^\nu\ =\ 1\quad\hbox{for every}\;\nu\in (0,1)\;.
$$
For a large set of potentials $V$ with singularities not stronger than Coulombic ones, more precisely, for all those satisfying : 
\bq\label{V1-V2} 
\lim_{|x|\to +\infty}V(x)=0\quad\mbox{and}\quad -\frac{\nu}{|x|}-c_1\leq V\leq c_2=\Sup(V)\;,
\eq
with $\nu\in (0,1)$, $c_1$, $c_2\in\R$, $c_1$, $c_2\geq 0$, $c_1+c_2-1< \sqrt{1-\nu^2}$, the following result was proved in \cite{Dolbeault-Esteban-Sere-00B} :
\begin{theorem}\label{TR5} {\rm\cite{Dolbeault-Esteban-Sere-00B}} Let $V$ be a radially symmetric function satisfying (\ref{V1-V2}) and define $\lambda_1 (V)$ as the smallest eigenvalue of $H_0 + V$ in the interval $(-1, 1)$. Then for all $\varphi \in H^1 (\R^3, \C^2)$,
\bq\label{R5} 
\Int_{\R^3}\left(\Frac{|{\boldsymbol{\sigma}}\cdot{\boldsymbol{\nabla}}\varphi|^2}{1+\lambda_1(V)-V}\ +\ \Big(1-\lambda_1(V)\Big)\,|\varphi|^2\right)\,dx\ \geq\ -\Int_{\R^3}V\,|\varphi|^2\,dx\;.
\eq\end{theorem}
Inequality (\ref{R5}) is achieved by the {\sl large component\/}, {\sl i.e.\/}, the two-spinor made of the first two complex valued components of the four-spinor, of any eigenfunction associated with $\lambda_1(V)$. In particular if $V=-\frac\nu{|x|}$, $\nu\in(0,1)$, we get
\begin{corollary} \label{CR6} {\rm\cite{Dolbeault-Esteban-Sere-00B}} For any $\nu\in(0,1)$, for all $\varphi\in H^1(\R^3,\C^2)$,
\bq\label{R6}
\Int_{\R^3}\left(\Frac{|{\boldsymbol{\sigma}}\cdot{\boldsymbol{\nabla}}\varphi|^2}{1+
\sqrt{1-\nu^2}+\frac{\nu}{|x|}}\ +\ \Big(1-\sqrt{1-\nu^2}\Big)\ |\varphi|^2\right)\,dx\ \geq\ \nu\ \Int_{\R^3}\Frac{|\varphi|^2}{|x|}\,dx\;.
\eq
\end{corollary}
This inequality is achieved in $L^2 \left(\R^3,|x|^{-1}dx\right)^4$. Inequality (\ref{R1}) is obtained from~(\ref{R6}) by taking the limit $\nu \rightarrow 1$. Theorem \ref{TR1} is therefore a straightforward consequence of Corollary \ref{CR6}. Note that (\ref{R1}) is not achieved in $L^2\left(\R^3,|x|^{-1}dx\right)^4$.

\medskip The aim of this paper is twofold. On the one hand, we give a direct analytical proof of Theorem \ref{TR1} which does not use any {\sl a priori }spectral knowledge on the operator $H_\nu$. On the other hand, we prove more general inequalities by showing that for some continuous functions $W>1$ a.e. and constants $R>0$ and $C(R)\leq 0$, the inequality
\bq\label{R8} 
\Int_{\R^3}\left(\Frac{|{\boldsymbol{\sigma}}\cdot{\boldsymbol{\nabla}}\varphi|^2}{1+\frac{W(|x|)}{|x|}}\ +\ |\varphi|^2\right)\,dx\ \geq\ \Int_{\R^3}\Frac{W(|x|)}{|x|}\,|\varphi|^2\,dx\ +\ C(R)\Int_{S_R}|\varphi|^2\,d\mu_R
\eq
holds for all $\varphi\in H^1(\R^3,\C^2)$. Here $\mu_R$ is the surface measure induced by Lebesgue's measure on the sphere $S_R:=\{x\in\R^3\;:\; |x|=R\}$. Note that this inequality is relevant for the Dirac operator with potential \hbox{$V=\frac W{|x|}$}. Improved inequalities like (\ref{R1}) or (\ref{R8}) with the operator ${\boldsymbol{\sigma}}\cdot{\boldsymbol{\nabla}}$ replaced by ${\boldsymbol{\nabla}}$ can easily be obtained by considering separately the real and the imaginary parts of the components of the two-spinors.

We are actually interested in understanding for which functions $W$ Inequality~(\ref{R8}) holds and what is the {\sl optimal\/} behavior of the function $W$ near $0$ or near $+ \infty$. By optimal at $s=0$ or $s=+\infty$, we mean optimal at each order in the sense that we look for a expansion of the form $W=1+\sum_{k=1}^{+\infty}c_k\,W_k$ such that at each order $k_0$, $0\leq W_{k_0+1}=o(W-1-\sum_{k=1}^{k_0}c_k\,W_k)$ and $c_{k_0+1}$ is the largest possible constant. What we are going to prove is a little bit involved: it is not clear that $c_k$ is a constant (see Appendix C) and we are only going to prove that the maximum of its $\liminf$ is achieved. As in Hardy-like inequalities for the Laplacian operator on balls centered at the origin, we will see that the optimal behavior of the function $W$ near $0$ is a logarithmic perturbation of the constant $1$. More precisely the optimal behavior near the origin is given by functions of the form 
$$
{W_\infty}(x)\ =\ 1\ +\ \Frac{1}{8}\,\Sum^\infty_{k=1}\,X_1(|x|)^2\cdots X_k(|x|)^2\;, 
$$
where $X_1(s):=(a-\log(s))^{-1}$ for some $a>1$, $X_k(s):= X_1\circ X_{k-1})$. The functions $X_k$ and $W^\infty$ are well defined for $|x|=s<e^{a-1}$ (see Appendix A  for basic properties of the functions $X_k$). These asymptotics are optimal in the above sense, with $W_k=X_1(|x|)^2\cdots X_k(|x|)^2$ and $\liminf c_k=\frac 18$. On the other hand, as $|x|\to + \infty$, the optimal growth for $W$ is given by $|x|$, {\sl i.e.\/}, the first term in the l.h.s. of (\ref{R8}) does not help.
\begin{theorem}\label{TR8} Assume that for some $R>0$ Inequality (\ref{R8}) holds for every spinor $\varphi\in C_0^{\infty}(\R^3,\C^2)$, where $W$ is a radially symmetric continuous function from $\R^+$ to $\R^+$. Assume moreover that $W(0)>0$ and $W$ is nondecreasing in a neighbourhood of $0^+$. Then $W(0)\leq 1$, 
$$
\limsup_{s\to +\infty}W(s)/s\leq 1
$$
and for all $k\geq 1$, 
\bq\label{twentytwo}
\liminf_{s\to 0^+}\Big(W(s)-1-\frac{1}8\Sum_{j=1}^k X_1^2(s)\cdots X_j^2(s)\Big)X_1^{-2}(s)\cdots X_{k+1}^{-2}(s) \leq\frac{1}{8}\;.
\eq
As soon as $W\not\equiv 1$, $C(R)$ must be negative. Moreover, there are continuous functions $W \geq 1$ for which (\ref{R8}) holds with some $R>0$, $C(R)<0$, such that $\lim_{s \to +\infty} W(s)/s = 1$ and \eqn{twentytwo} holds with equality for all $k\geq 1$. \end{theorem}
Note that this result is independent of the particular value of $a>1$ which appears in the definition of the functions $X_k$. The fact that we have to introduce a discontinuity at some $s=R$ is not contradictory with the known results for the usual Hardy inequality, for which bounded domains are considered and $q$ is taken large enough.

In Section \ref{S2} we give a direct analytical proof of Corollary \ref{CR6} together with several auxiliary results. We recall that Theorem \ref{TR1} is a straightforward consequence of Corollary \ref{CR6}. Section \ref{S3} is devoted to the proof of Theorem \ref{TR8}. Properties of the functions $X_k$, an existence result for a singular ODE needed for Theorem \ref{TR8} and an example illustrating why we have to consider a $\liminf$ in this theorem are given in three appendices.

\section{Proof of Theorem \ref{TR1}}\label{S2}
In this Section, we actually prove Corollary \ref{CR6}, which is slightly more than Theorem \ref{TR1}.

First we fix some notations. The spinor $\varphi=\left(\begin{smallmatrix}\varphi_1\\ \varphi_2\end{smallmatrix}\right)$ takes its values in $\C^2$ and by $|\varphi|^2$, $|{\boldsymbol{\nabla}}\varphi|^2$ and $|{\boldsymbol{\sigma}}\cdot{\boldsymbol{\nabla}}\varphi|^2$ we denote, respectively, the quantities $|\varphi_1|^2+|\varphi_2|^2$, $\Sigma_{k=1}^3(|\partial_k\varphi_1|^2+|\partial_k\varphi_2|^2)$ and $|\partial_3\varphi_1+\partial_1\varphi_2-i\,\partial_2\varphi_2|^2+|\partial_1\varphi_1+i\,\partial_2\varphi_1-\partial_3\varphi_2|^2$. Further, we notice that the Pauli matrices are Hermitian and satisfy the following properties :
$$
\sigma_j\,\sigma_k+\sigma_k\,\sigma_j=2\,\delta_{jk}\,\Id\,,\quad\forall\;j,\,k=1,\,2,\,3\;.
$$
With a standard abuse of notations, each time a scalar $\delta$ appears in an identity involving operators acting on two-spinors, it has to be understood as $\delta\,\Id$, where $\Id$ is the identity operator.

On the other hand, for all vectors ${\boldsymbol{a}},{\boldsymbol{b}}\in\C^3$, we have
$$
({\boldsymbol{\sigma}}\cdot{\boldsymbol{a}})({\boldsymbol{\sigma}}\cdot{\boldsymbol{b}}) ={\boldsymbol{a}}\cdot{\boldsymbol{b}}+i\,{\boldsymbol{\sigma}}\cdot({\boldsymbol{a}}\times{\boldsymbol{b}})\;.
$$
Applying this formula to ${\boldsymbol{a}}={\boldsymbol{x}}$ and ${\boldsymbol{b}}=i{\boldsymbol{\nabla}}$, we obtain the following expression for the commutator of ${\boldsymbol{\sigma}}\cdot{\boldsymbol{\nabla}}$ and ${\boldsymbol{\sigma}}\cdot{\boldsymbol{x}}$ :
$$
[{\boldsymbol{\sigma}}\cdot{\boldsymbol{\nabla}},\ {\boldsymbol{\sigma}}\cdot{\boldsymbol{x}}]\ =\ {\boldsymbol{\nabla}}\cdot{\boldsymbol{x}}-{\boldsymbol{x}}
\cdot{\boldsymbol{\nabla}}+2\,{\boldsymbol{\sigma}}\cdot{\boldsymbol{L}}=3+2\,{\boldsymbol{\sigma}}\cdot{\boldsymbol{L}}\;, 
$$
where ${\boldsymbol{L}}\ =\ -i\,{\boldsymbol{x}}\wedge{\boldsymbol{\nabla}}$ is the orbital angular momentum operator. The main point to note here is that ${\boldsymbol{L}}$ acts only on the angular variables.

For simplicity, for any function $h:\,\R^+\rightarrow\R$, we denote the functions $x\mapsto h(|x|)$ and $x\mapsto h'(|x|)$ by $h$ and $h'$ respectively. Now if such a function $h$ is differentiable a.e. in $\R^+$ and continuous in $[0,R)\cup(R,+\infty)$, we have
$$\label{C}\begin{array}{rl}\ds
[{\boldsymbol{\sigma}}\cdot{\boldsymbol{\nabla}},\ ({\boldsymbol{\sigma}}\cdot{\boldsymbol{x}})\ h]& =\ds\ |{\boldsymbol{x}}|\ h'
+(3+2\,{\boldsymbol{\sigma}}\cdot{\boldsymbol{L}})\ h+R\,[h]_R\,\delta_R\\
&\ds=\ 2\,(1+{\boldsymbol{\sigma}}\cdot{\boldsymbol{L}})\ h+h+|{\boldsymbol{x}}|\ h'+R\,[h]_R\,\delta_R\;,
\end{array}$$
where by $[h]_R:=h(R_+)-h(R_-)$ we denote the possible jump of $h$ at $R$ and $\delta_R$ is the Dirac delta function at $r=R$, in spherical coordinates.

The spectrum of the operator $1+{\boldsymbol{\sigma}}\cdot{\boldsymbol{L}}$ is the discrete set $\{\pm 1,\pm 2,\cdots\}$ (see~\cite{Thaller-92}). This can be seen by noticing that
$$
1+{\boldsymbol{\sigma}}\cdot{\boldsymbol{L}}={\boldsymbol{J}}^2-{\boldsymbol{L}}^2+\frac{1}4\,,\quad{\boldsymbol{J}}={\boldsymbol{L}}+\frac{{\boldsymbol{\sigma}}}2\;.
$$
Then, the fact that the spectrum of ${\boldsymbol{J}}^2$ (resp. ${\boldsymbol{L}}^2$) is the set $\{j(j+1)\,;\; j=\frac{1}2, \frac{3}2, \dots\}$ (resp. $\{\ell(\ell+1)\,;\;\ell=j\pm\frac{1}2,\;j=\frac{1}2,\,\frac{3}2,\,\dots\}$ proves the above result. The main point here is that $0$ is not in the spectrum of $1+{\boldsymbol{\sigma}}\cdot{\boldsymbol{L}}$. If we denote by $X_+$ (resp. $X_-$) the positive (resp. negative) spectral space of $1+{\boldsymbol{\sigma}}\cdot{\boldsymbol{L}},$ and by $P_\pm=\frac{1}2\big(1\pm\frac{1+{\boldsymbol{\sigma}}\cdot{\boldsymbol{L}}}{|1+{\boldsymbol{\sigma}}\cdot{\boldsymbol{L}}|}\big)$ the corresponding projectors on $H^1(\R^3,\C^2)$, for all $\varphi\in H^1(\R^3,\C^2)$, for all $h$ as above,
$$
(\varphi_+,[{\boldsymbol{\sigma}}\cdot{\boldsymbol{\nabla}},({\boldsymbol{\sigma}}\cdot{\boldsymbol{x}})\,h]\,\varphi_+\!)\geq\Int_{\R^3}\!(3h+|{\boldsymbol{x}}|\,h')|\varphi_+|^2\,dx+R\,[h]_R\!\Int_{S_R}\!|\varphi_+|^2\d\mu_R\;, 
$$
$$
(\varphi_-,[{\boldsymbol{\sigma}}\cdot{\boldsymbol{\nabla}},({\boldsymbol{\sigma}}\cdot{\boldsymbol{x}})\,h]\,\varphi_-\!)\leq\Int_{\R^3}\!(-h+|{\boldsymbol{x}}|\,h')|\varphi_-|^2\,dx+R\,[h]_R\!\Int_{S_R}\!|\varphi_-|^2\d\mu_R\;, 
$$
where $\varphi_\pm :=P_\pm\varphi$.

By Cauchy-Schwartz' inequality, for any measurable function $g:\R^+\rightarrow \R^+$,
\begin{eqnarray} 
\Int_{\R^3}\Big(3h+|{\boldsymbol{x}}|\,h'\Big)|\varphi_+|^2\,dx&\leq&\Int_{\R^3}g\,|({\boldsymbol{\sigma}}\cdot{\boldsymbol{\nabla}})\varphi_+|^2\,dx\nonumber\\
\label{C+}\\
&&+\Int_{\R^3}\Frac{|x|^2\,h^2}{g}\,|\varphi_+|^2\,dx-R\,[h]_R\int_{S_R}|\varphi_+|^2\,d\mu_R\;,\nonumber\\
\nonumber\end{eqnarray}
where again we abbreviate $g(|x|)$ by $g$.

\medskip Define now $W$ and $m:\R^+\rightarrow\R$ by
$$
g(s)\ =\ \Frac{s}{W(s)+s}\quad{\rm and}\quad m(s)\ =\ s\,h(s)\;,
$$
and assume that $W$ is positive on $\R^+$. With the same notation as above, we can rewrite (\ref{C+}) as
\begin{eqnarray}
&\Int_{\R^3}\frac{1}{s}\,\big(2m+s\,m'-(s+W)\,m^2\big)_{\vert s=|x|}|\varphi_+|^2\,dx\nonumber\\
\label{I+}\\
&\qquad\qquad\qquad\qquad\qquad\leq\ \Int_{\R^3}\Frac{|({\boldsymbol{\sigma}}\cdot{\boldsymbol{\nabla}})\varphi_+|^2}{1\ +\ \frac{W}{|x|}}\,dx-R\,[h]_R\!\Int_{S_R}\!|\varphi_+|^2\,d\mu_R\;.\nonumber
\end{eqnarray}

Similarly, for $\varphi_-:=P_-\varphi$, we find
\begin{eqnarray}
&\Int_{\R^3}\frac{1}{s}\,\big(2m-s\,m'-(s+W)\,m^2\big)_{\vert s=|x|}|\varphi_-|^2\,dx\nonumber\\
\label{I-}\\
&\qquad\qquad\qquad\qquad\qquad\leq\ \Int_{\R^3}\Frac{|({\boldsymbol{\sigma}}\cdot{\boldsymbol{\nabla}})\varphi_-|^2}{1\ +\ \frac{W}{|x|}}\,dx+R\,[h]_R\!\Int_{S_R}\!|\varphi_-|^2\,d\mu_R\;.\nonumber
\end{eqnarray}

Note that for any measurable radial function $b$, the spaces $X_+$ and $X_-$ are also orthogonal in $L^2(\R^3;b(|x|)\,dx)$. Moreover, we have
\begin{lemma}$P_-\big({\boldsymbol{\sigma}}\cdot{\boldsymbol{\nabla}}\big)^2P_+\ \equiv\ P_+\big({\boldsymbol{\sigma}}\cdot{\boldsymbol{\nabla}}\big)^2P_-\ \equiv\ 0\quad{\rm in}\;H^1(\R^3,\C^2)$.\end{lemma}
\proof A direct computation shows that the anti-commutator $\{{\boldsymbol{\sigma}}\cdot{\boldsymbol{\nabla}},1+ {\boldsymbol{\sigma}}\cdot{\boldsymbol{L}}\}= 0$, {\sl i.e.\/}, ${\boldsymbol{\sigma}}\cdot{\boldsymbol{\nabla}}$ anticommutes with $1+{\boldsymbol{\sigma}}\cdot{\boldsymbol{L}}$. Hence $({\boldsymbol{\sigma}}\cdot{\boldsymbol{\nabla}})^2$ commutes with $1+{\boldsymbol{\sigma}}\cdot{\boldsymbol{L}}$. 

Now, let $\Phi_\pm\in X_\pm$ be two eigenfunctions of $1+{\boldsymbol{\sigma}}\cdot{\boldsymbol{L}}$ with eigenvalues $\lambda_\pm$, $\lambda_-<0<\lambda_+$. Then,\begin{eqnarray*}
\left(({\boldsymbol{\sigma}}\cdot{\boldsymbol{\nabla}})\Phi_-, ({\boldsymbol{\sigma}}\cdot{\boldsymbol{\nabla}})\Phi_+\right)&=&\frac{1}{\lambda_+}\left(\Phi_-, ({\boldsymbol{\sigma}}\cdot{\boldsymbol{\nabla}})^2(1+{\boldsymbol{\sigma}}\cdot {\boldsymbol{L}})\Phi_+\right)\\
&=&\frac{1}{\lambda_+}\left(\Phi_-, (1+{\boldsymbol{\sigma}}\cdot {\boldsymbol{L}})({\boldsymbol{\sigma}}\cdot{\boldsymbol{\nabla}})^2\Phi_+\right)\\
&=& \frac{1}{\lambda_+}\left((1+{\boldsymbol{\sigma}}\cdot {\boldsymbol{L}})\Phi_-, ({\boldsymbol{\sigma}}\cdot{\boldsymbol{\nabla}})^2\Phi_+\right)\\
&=& \frac{\lambda_-}{\lambda_+}\left(\Phi_-, ({\boldsymbol{\sigma}}\cdot{\boldsymbol{\nabla}})^2\Phi_+\right)\;,
\end{eqnarray*}
which is impossible except if $ (({\boldsymbol{\sigma}}\cdot{\boldsymbol{\nabla}})\Phi_-, ({\boldsymbol{\sigma}}\cdot{\boldsymbol{\nabla}})\Phi_+)=0$. \finprf

Adding (\ref{I+}) and (\ref{I-}), we get the following result.
\begin{proposition} \label{PP1} Let $W$ be a positive measurable function on $\R^+$ and consider two functions $m_\pm : \R^+\to\R$ such that the maps $s\mapsto m_\pm(s)/s$ are continuous on $ [0,R)\cup (R,+\infty)$ and differentiable a.e. on $\R^+$. Then for any $\varphi\in H^1(\R^3,\C^2)$, 
\begin{eqnarray*}\label{R9}
&&\Int_{\R^3}\Frac{|({\boldsymbol{\sigma}}\cdot{\boldsymbol{\nabla}})\varphi|^2}{1\ +\ \frac{W}{|x|}}\,dx\,\mp\,\Sum_\pm\; [m_\pm]_{_R}\,\Int_{S_R}\!|P_\pm\varphi|^2\,d\mu_R\nonumber\\
&&\qquad\geq\;\Sum_\pm\Int_{\R^3}\frac{1}{s}\,\big(2m_\pm\pm s\,m_\pm'-(s+W)\,m_\pm^2\big)_{\vert s=|x|}|P_\pm\varphi|^2\,dx\;.\\
\end{eqnarray*}
\end{proposition}
In order to prove Inequality (\ref{R8}), we have to find two functions, $m_+$ and $m_-$, and a continuous function $W\geq 1$ such that for $ s\in\R^+$ a.e.,
$$
2m_{\pm}\pm s\,m'_{\pm}-(W+s)\,m^2_{\pm}\ \geq\ W-s\;.\label{M}
$$
This means
\bq
1\leq W\ \leq\ \Frac{2m_\pm\pm s\,m'_\pm-s\,m^2_\pm +s}{1+m^2_\pm}\label{W}\;.
\eq
Moreover $W$ has to be as large as possible (near the origin and near infinity), {\sl i.e.\/} {\sl optimal\/} in the sense of Section \ref{S1}. Then (\ref{R8}) follows with $C(R)=-\max([m_-]_R,[m_+]_R)$.

In the sequel, for every function $m$ as in Proposition \ref{PP1}, we will use the notation
\bq\label{Wm}
W^{\pm,m}:=\Frac{2m\pm s\,m'-s\,m^2+s}{1+m^2}\;.
\eq

\medskip\noindent{\sl Proof of Theorem \ref{TR1} and Corollary \ref{CR6}.\/} 
This is simply done by choosing $m_+\equiv m_-\equiv 1$ or $m_+\equiv m_-\equiv\frac{1-\sqrt{1-\nu^2}}{\nu}$ in Proposition \ref{PP1}. Since in both cases the functions $s\to m_\pm\,/s$ are continous in~$\R^+$, there is no surface integral term in those inequalities : $C(R)=0$. \finprf

\begin{remark} The above arguments leading to the proof of Theorem \ref{TR1} can also be viewed as a {\rm completing the square} strategy : for all $\varphi\in H^1(\R^3)$, $h$, $g\in C(\R^+,\R^+)$, $h$ differentiable a.e. in $\R^+$, it is clear that
$$ 
\Int_{\R^3}\Big\vert\,\sqrt{g}\ ({\boldsymbol{\sigma}}\cdot{\boldsymbol{\nabla}})P_\pm\varphi\ \pm\ \Frac{({\boldsymbol{\sigma}}\cdot{\boldsymbol{x}})\,h}{\sqrt{g}}\ P_\pm\varphi\,\Big\vert^2\,dx\ \geq\ 0\;. 
$$
Expanding the squares in the above expressions, integrating by parts the cross terms and adding the two inequalities, we find (\ref{R8}) if $W$ satisfies (\ref{W}). 
\end{remark}

\section{Proof of Theorem \ref{TR8}}\label{S3}

\subsection{Direct estimates}\label{S3-1}

We start this section by finding the optimal behavior near the origin and near infinity for continuous functions $W$ for which (\ref{R8}) holds, where $W$ is given or not by some function $m$ as defined in (\ref{Wm}). This is done in a series of intermediate results.
\begin{lemma} \label{PP2} Let $W$ be any function satisfying $W\geq 1$ on $\R^+$ and for which Inequality (\ref{R8}) holds. Then, necessarily, 
$$
\lim_{s\to 0^+}\,W(s)=1\quad\mbox{and}\quad\limsup_{s\to +\infty}\,\frac{W(s)}{s}\leq 1\;.
$$
\end{lemma}
\proof By assumption, $W\geq 1$. If we had $W(0)>1$, it would be easy to contradict the fact that $1$ is the best constant in \eqn{R2} and \eqn{R3}. As $s\to+\infty$, the result follows from the simple observation that by scaling we can easily construct functions $\varphi_n:=n^{-3/2}\varphi(\cdot/n)$ such that
$$ 
\int_{\R^3}|\varphi_n|^2\,dx=1,\quad\int_{\R^3}|{\boldsymbol{\sigma}}\cdot{\boldsymbol{\nabla}}\varphi_n|^2\,dx\to 0\quad\hbox{as}\quad n\to +\infty\;,
$$
so that the gradient term does not play any role.\finprf

\begin{proposition}\label{PP3} Let $m_\pm\in C([0,\delta))$ for some $\delta>0$. Consider $W^{\pm,m_\pm}$ defined according to \eqn{Wm} and let $W:=\min(W^{+,m_+},\,W^{-,m_-})$ be a function for which $W\geq 1$. Then (\ref{R8}) holds, $m_\pm(0)=1$, $m_\pm\geq 1$ in a neighbourhood of $s=0^+$ and 
\bq
\limsup_{s\to 0^+}\,\big|(m_\pm(s)-1)\log(s)\big|<+\infty\;.
\eq
\end{proposition} 
\proof We prove this for $m_+$, the proof for $m_-$ being identical. Let us write $m_+ = n +1$. Then (\ref{W}) is equivalent to
\bq\label{WO}
0\ \leq\ W-1=\Frac{s\,n'-n^2-2\,s\,n-s\,n^2}{2+2n+n^2}\;.
\eq
From this inequality, we infer that $n'\geq n(n+2)+n^2/s$, so that there are two possibilities for the behavior of $n$ near $0$ : either $n$ is monotone and $\lim_{s\rightarrow 0^+}n(s)=a\in(-\infty,+\infty]$, or $n$ oscillates near $0$ in the interval $(-2,0)$. The latter case is impossible because on the sequence of local minima approaching $0$, the r.h.s. of (\ref{WO}) would eventually be negative. So $\lim_{s\rightarrow 0^+}n(s)=a\in(-\infty,+\infty]$ and if $a\ne 0$, for $s>0$ small,
$$
s\,n'\sim n^2\;,
$$
which by integration implies that near $0$,
\bq
n(s)\sim\Frac{1}{|\log s|+C}\;,\quad C\geq 0\label{N0} 
\eq
for some constant $C\geq 0$, a contradiction. Hence, necessarily, $a=0$ and the result follows from \eqn{N0}, which still holds true when $a=0$.\finprf

Next we prove the following asymptotic result :
\begin{lemma}\label{cascade} Let $\mathcal{A}$ denote the class of the functions $n$, continuous in the interval $[0,\delta)$ for some $\delta>0$, and such that $n(0)=0$. Then, for all $k\geq 1$,
$$
\label{optimm} \sup_{n\in\mathcal{A}}\left\{\liminf_{s\to 0^+}\Big(s\,n'(s)\!-\!n^2(s)\!-\!\frac{1}{4}\Sum_{j= 1}^{k-1}X_1^2(s)\cdots X_j^2(s)\Big)X_1^{-2}(s)\cdots X_{k}^{-2}(s)\,\right\}\!=\!\frac{1}{4}\,.
$$
\end{lemma}
The fact that we are dealing with a $\liminf$ and not a $\limsup$ may look surprising at first sight. However, an upper limit cannot be expected, as it is shown by the example given at the end of the paper.

\proof For $0\leq s<1$ and $X_1(s):=(a-\log s)^{-1}$, define implicitely $n_1$ by
$$
n(s)=\frac 1{2}\,X_1(s)\,\left( 1-2\,n_1\left(\frac 1{X_1(s)}\right)\right)\;.
$$
Then
$$
s\,n'(s)-n^2(s)=\frac1{4\,t^2}+\frac{\,t\,n_1'(t)-n_1^2(t)}{t^2}\,,\quad t=\frac 1{X_1(s)}\in(1,+\infty)\;.
$$
Next, for all $k\geq1$ and $s>1$, let us define again $n_{k+1}$ in terms of $n_k$ by 
$$
n_k(s):=\frac{1}{2\,t}\,\Big(2\,n_{k+1}(t)-1 \Big)\,,\quad t=\frac 1{X_1(1/s)}\in(1,+\infty)\;.
$$
Then
$$
s\,n_k'(s)-n_k^2(s)= \frac1{4\,t^2}+ \frac{t\, n_{k+1}'(t)-n_{k+1}^2(t) }{t^2}\;.
$$
Hence, for every $k\geq1$ and every $0\leq s<1$, with $z=1/X_{k}(s)$, we have
$$
s\,n'(s)-n^2(s)=\frac{1}{4} \Sum_{j=1}^k X_1(s)^2\cdots X_j(s)^2+\,X_1(s)^2\cdots X_{k}(s)^2\left({z\, n_{k}'(z)-n_{k}^2(z) }\right)\;.
$$

\smallskip\noindent 1) Choosing $n_{k}=0$ delivers a function $n(s)$ with
$$
s\,n'(s)-n^2(s)=\frac{1}{4}\Sum_{j=1}^k X_1(s)^2\cdots X_j(s)^2\;.
$$
Note that in this case, $n(s)=\sum_{j=1}^k X_1(s)\cdots X_j(s)$ (see Appendix A for more details). This shows that 
\bq
\sup_{\mathcal{A}}\left\{\liminf_{s\to 0^+}\Big(s\,n'(s)\!-\!n^2(s)\!-\!\frac{1}{4}\Sum_{j=1}^{k-1}X_1^2(s)\cdots X_j^2(s)\Big)X_1^{-2}(s)\cdots X_{k}^{-2}(s)\right\}\!\geq\! \frac{1}{4}\;. 
\eq

\smallskip\noindent 2) Let now $n$ be any function in $\cal A$. For every $k$, 
$$
\liminf_{t\to+\infty}\; (t\,n'_k-n_k^2)\leq 0\;.
$$
If the above limit was to be larger than $0$, say some constant $b>0$, then, integrating the inequality $t\,n_k'\geq n_k^2$ would show that $n_k$ tends to $0$ at infinity, while on the other hand, integrating $ t\,n_k'\geq b/2$ would show that $n_k$ is unbounded near infinity, which provides an obvious contradiction.\finprf
\begin{corollary}\label{CC1} Let $W$ be as in Proposition \ref{PP3}. Then (\ref{twentytwo}) holds and the optimal asymptotic behavior near the origin is achieved.\end{corollary}
\proof Close to $s=0^+$, the fact that
$$
\Frac{s\,n'-n^2-2\,s\,n-s\,n^2}{2+2n+n^2}\sim\frac 12\,(s\,n'-n^2)
$$
immediately provides
$$
\liminf_{s\to 0^+}\Big(W(s)-1-\frac{1}8\Sum_{j=1}^k X_1^2(s)\cdots X_j^2(s)\Big)X_1^{-2}(s)\cdots X_{k+1}^{-2}(s) \leq\frac{1}{8}
$$
and the optimal behaviour is achieved, for instance, by 
$$
W=\bar W:=\min(W^{+,1+\bar n}, W^{-,1-\bar n})\;,\quad \bar n:=\frac{1}2 \Sum_{j=1}^{+\infty} X_1(s)\cdots X_j(s)\;,
$$
so that 
$$
{\bar W(s)}= 1+\frac{1}{8}\Sum_{j=1}^{+\infty} X_1^2(s)\cdots X_j^2(s)+o(s)
$$
(see Appendix A for more details).\finprf

\subsection{Estimates based on improved Hardy inequalities for the Laplacian}\label{S3-2}

The above arguments show that the optimal growth near $0$ and near infinity for any function $W$ generated (as above) by functions $m_\pm$, continuous near the origin and near infinity, and for which (\ref{R8}) holds, is given by (\ref{twentytwo}) with equality for each $k\geq 1$, as in the statement of Theorem \ref{TR8}. On the other hand, the optimality near infinity was established in Lemma \ref{PP2}. However, it remains to prove that there is no function $W$ -- not given by \eqn{Wm} -- with higher growth at the origin. This amounts to prove that there is no radial function $W$ with more singular asymptotics near the origin and for which the differential problem
$$
s\,m'=s\,m^2-s-2m+(1+m^2)\,W\;,\quad m(0)=1\;,
$$
cannot be solved for some function $m$, continuous at $0$. The rest of this section is devoted to this question.

\medskip\noindent\underline{\sl Step 1 :\/} We first remark that in this problem the angular variables do not play any role : only radially symmetric spinors of a particular form are relevant to obtain the optimal asymptotics. 
\begin{proposition}\label{Prop3} Let $W:\R^+\to \R^+$ be a radially symmetric continuous and a.e. differentiable function. Assume that a.e. $r\in [0,R)$, 
\bq\label{rpetit}
-r\leq r\,W'(r)\leq 3\,W+2\,r\;.
\eq
Then, for all $\varphi \in H^1_0 (B_R,\C^2)$,
$$
\int_{\R^3} \frac{r}{(r+W)} \;|{\boldsymbol{\sigma}}\cdot{\boldsymbol{\nabla}} \varphi|^2\,dx\geq \int_{\R^3} \frac{r}{(r+W)}\,|\partial_r\varphi|^2\,dx\;,
$$
and the optimizers are radially symmetric and of the form $\varphi=\left(\begin{smallmatrix} v(r) \cr 0 \end{smallmatrix}\right)$.  In particular, if $W(0)>0$ and $W$ is nondecreasing near $0$, \eqn{rpetit} holds true for $R$ sufficiently small. \end{proposition}
\proof Let $r=|x|$. By $\partial_r$, we mean ${\boldsymbol{\nabla}}\cdot\frac{\boldsymbol{x}}r$. For all $2$-spinor $\varphi$ with compact support in the ball $B_R$, using $(\boldsymbol{\sigma}\cdot\frac{\boldsymbol{x}}r)^2=1$, we have
\begin{eqnarray*}
&&\hspace*{-0.5cm}\int_{\R^3} \frac{r}{(r+W)} \;|{\boldsymbol{\sigma}}\cdot{\boldsymbol{\nabla}} \varphi|^2\,dx\\
&=&\int_{\R^3}\frac{r}{(r+W)}\;\left|({\boldsymbol{\sigma}}\cdot \frac{{\boldsymbol{x}}}r)({\boldsymbol{\sigma}}\cdot{\boldsymbol{\nabla}})\varphi\right|^2\,dx\\
&=&\int_{\R^3} \frac{r}{(r+W)}\;\left|\partial_r\varphi-\frac{1}{r}{\boldsymbol{\sigma}}\cdot {\boldsymbol{L}}\varphi\right|^2\,dx\\
&=&\int_{\R^3} \frac{r}{(r+W)}\,\Big(|\partial_r\varphi|^2+\frac{1}{r^2}|{\boldsymbol{\sigma}}\cdot {\boldsymbol{L}}\varphi|^2\Big)\,dx\\
&&\hspace*{2cm} -\int_0^{+\infty}\frac{r^2}{(r+W)}\,\partial_r\left(\int_{S^2}<\varphi, {\boldsymbol{\sigma}}\cdot {\boldsymbol{L}}\varphi>\right)\,dr\\
&=&\int_{\R^3} \frac{r}{(r+W)}\,\Big(|\partial_r\varphi|^2+\frac{1}{r^2}|{\boldsymbol{\sigma}}\cdot {\boldsymbol{L}}\varphi|^2\Big)\,dx \\
&&\hspace*{2cm} +\int_0^{+\infty}\partial_r\,\Big(\frac{r^2}{(r+W)} \Big)\, \left(\int_{S^2}<\varphi, {\boldsymbol{\sigma}}\cdot {\boldsymbol{L}}\varphi>\right)\,dr\;.
\end{eqnarray*}
Now, if we choose $\varphi$ belonging to the class of spinors generated by the eigenfunctions of ${\boldsymbol{\sigma}}\cdot {\boldsymbol{L}}$ with eigenvalue $n$, we notice that
\begin{eqnarray*}
&&\hspace*{-0.5cm}\int_{\R^3} \frac{r}{(r+W)} \;|{\boldsymbol{\sigma}}\cdot{\boldsymbol{\nabla}} \varphi|^2\,dx\\
&=&\int_{\R^3} \frac{r}{(r+W)}\,|\partial_r\varphi|^2\,dx +\left(\frac{n^2}{r\,(r+W)} +\frac{n}{r^2}\Big(\frac{r^2}{r+W} \Big)'\right)\,|\varphi|^2\,dx\\
&=&\int_{\R^3} \frac{r}{(r+W)}\,|\partial_r\varphi|^2\,dx
+\left(\frac{(n^2+2n)(r+W)-(1+W')\,r\,n} {r\,(r+W)^2} \right)\,|\varphi|^2\,dx\;,
\end{eqnarray*}
which implies that for all $\varphi$ supported in $B_R$, 
\bq\label{radineq}
\int_{\R^3} \frac{r}{(r+W)} \;|{\boldsymbol{\sigma}}\cdot{\boldsymbol{\nabla}} \varphi|^2\,dx\geq \int_{\R^3} \frac{r}{(r+W)}\,|\partial_r\varphi|^2\,dx\;,
\eq
and the optimizers for this inequality are radially symmetric.

Indeed, remember that the spectrum of $(1+{\boldsymbol{\sigma}}\cdot {\boldsymbol{L}})±,$ is the set $\{ \pm 1, \pm 2, \dots\}$. Hence, $n\in\{\dots, -3, -2, 0, 1, 2, \dots\}$ But our assumptions imply that the minimum of $(n^2+2n)(r+W)-(1+W')\,n\,r$ on $B_R$ is nonnegative for $n\ne 0$ and $0$ for $n=0.$ Hence, the optimizers for (\ref{radineq}) correspond to spinors which are eigenfunctions of $1+{\boldsymbol{\sigma}}\cdot {\boldsymbol{L}}$ with eigenvalue $1$ ($n=0$). These spinors are radially symmetric and their second component is equal to $0$ (see \cite{Thaller-92}).

The last assertion of the proposition trivially follows from the fact that $W$ having a finite limit at $0$, $\lim_{r\to 0^+}\,r\,W'(r)$ must be equal to $0$. \finprf

\medskip\noindent\underline{\sl Step 2 :\/} We prove a relation between Hardy-like inequalities for the Laplacian and for the Dirac operator in the radially symmetric case. 

Consider a function $W:\R^+\to \R^+$ such that $W/r^3$ is integrable at infinity and define a new variable 
\bq\label{y-r} 
y(r):=\frac{1}{\int_r^{+\infty}(s+W)\,s^{-3}\,ds}=\frac{r^2}{\int_1^{+\infty}(t\,r+W(t\,r))\,t^{-3}\,dt} \;.
\eq
Now, for any $u\in C^\infty_0(\R^+,\R)$, we define $q(y):=u(r)$, where $y$ and $r$ are related by the above change of variables. Then straightforward computations show that the inequalities
\bq\label{twentythree}
\int_0^{+\infty} \frac{r^3}{r+W}\,|u'|^2\,dr\geq \int_0^{+\infty} W\,r\,|u|^2\,dr
\eq
and 
\bq\label{AA2}
\int_{0}^{+\infty} y^2\,|q'|^2\,dy\geq \int_{0}^{+\infty} V\,|q|^2\,dy
\eq
are equivalent, with $ V$ given in terms of $r=r(y)$ by
\bq\label{variablechange}
V(y)=\frac{W(r)\,r^4}{y^2\,(W(r)+r)} = \frac{W(r)}{(r+W(r))}\left( \int_1^{+\infty}(t\,r+W(t\,r))\,t^{-3}dt \right)^2\;.
\eq
\begin{proposition} Let $W:\R^+\to \R^+$ be such that $W/r^3$ is integrable at infinity. Then Hardy-like inequalities (\ref{twentythree}) and (\ref{AA2}) are equivalent, with $W$ and $V$ related by (\ref{y-r}) and (\ref{variablechange}). \end{proposition}

\begin{remark}Note that when dealing with functions which are compactly supported in a fixed ball, the behavior of $W$ near infinity is irrelevant, since $W$ can be modified outside the ball, without changing the integrals in the above inequalities. In particular, this is the case when searching for the optimal asymptotics near the origin of the functions $W$ for which (\ref{R8}) holds. \end{remark}

\medskip\noindent\underline{\sl Step 3 :\/} Let us focus now on improved Hardy inequalities for the Laplacian. Compared with (\ref{twentythree}), Inequality (\ref{AA2}) is easier to deal with, because the potential appears only in the r.h.s. In \cite{Barbatis-Filippas-Tertikas-1,Barbatis-Filippas-Tertikas-2,Filippas-Tertikas-02} (see also \cite{Adimurthi,Adimurthi-Esteban-03,Adimurthi-Ramaswamy-Chaudhuri-01,Adimurthi-Sandeep-01,Sandeep-01} for related results) we find the following optimality result :
\begin{theorem}\label{terti} {\rm\cite{Filippas-Tertikas-02}} The optimal asymptotical behavior near the origin for potentials $V$ for which the Hardy-like inequality (\ref{AA2}) holds for all $q\in C^\infty_0(\R^3, \R)$ is given at each order by 
\bq\label{BBB}
V_\infty(s)=\frac{1}4 \Big( 1+\Sum_{j=1}^{+\infty} X_1^2(s)\cdots X_j^2(s) \Big)\;.
\eq
\end{theorem}
\noindent{\sl An elementary proof for Theorem \ref{terti} in the radially symmetric case.\/} For completion, let us give a simple proof of this result. This can be done by using the same kind of changes of variables as those used in the proof of Lemma \ref{cascade}.

Let $a>1$ be the constant which appears in the definition of $X_1$ and take $R<e^a$. For all $u\in H^1_0(B_R)$, for every $k\geq 1$, define the functions $g_k$ by
\begin{eqnarray*}
&&u(r)=\frac 1{\sqrt{r}}\,g_1\left(\frac 1{X_1(1/r)}\right)\\
&&g_k(s):=\sqrt s\,g_{k+1}(t)\;,\quad t=\frac 1{X_1(1/s)}
\end{eqnarray*} 

A simple computation shows that 
$$
\int_0^R r^2\,|u'|^2\,dr=\frac{1}4\int_0^R|u|^2\,dr+\int_{X_1^{-1}(R)}^{+\infty}|g'_1|^2\,dy\;.
$$ 
With the notation $t=t(s)=1/X_1(1/s)=a+\log s$, it is clear that $s\,\frac {dt}{ds}=1$. From the definition of $g_{k+1}$, we get, for any $k\geq 1$,
$$
g_k'(s)=\frac 1{2\sqrt s}\,g_{k+1}(t)+\sqrt s\,\frac {dt}{ds}\,g_{k+1}'(t)\;.
$$
Moreover, for any $A>0$,
\begin{eqnarray*}
&&\int_A^{+\infty}\left|\frac 1{\sqrt s}\,g_{k+1}(t(s))\right|^2\,ds=\int_{X_1^{-1}(A)}^{+\infty}\left|g_{k+1}(t)\right|^2\,dt\;,\\
&&\int_A^{+\infty}\left|\sqrt s\,\frac {dt}{ds}\,g_{k+1}'(t(s))\right|^2\,ds=\int_{X_1^{-1}(A)}^{+\infty}\left|g_{k+1}'(t)\right|^2\,dt\;.
\end{eqnarray*}
Taking $A>0$ small enough, this means 
$$
\int_A^{+\infty}\left|g_k'(s)\right|^2\,ds=\frac 14\int_{X_1^{-1}(A)}^{+\infty}\left|g_{k+1}(t)\right|^2\,dt +\int_{X_1^{-1}(A)}^{+\infty}\left|g_{k+1}'(t)\right|^2\,dt
$$
since $g_{k+1}$ has a compact support in $(0,+\infty)$. Thus
\bq\label{oppt}
\int_{X_k^{-1}(R)}^{+\infty}|g'_k|^2\,ds=\int_{X_{k+1}^{-1}(R)}^{+\infty}|g'_{k+1}|^2\,dt+ \frac{1}4\int_0^RX_1^2\cdots X_k^2\,|u|^2\,dr\;,
\eq
where by $X_{k+1}^{-1}$ we denote the inverse function of $X_{k+1}$,
$$
u=\Big(r\,X_1(r)\cdots X_k(r)\Big)^{-1/2}\,g_{k+1}\left(\frac 1{X_{k+1}(r)}\right)
$$
and
$$
\int_0^R r^2\,|u'|^2\,dr= \frac{1}4\int_0^R\Big(1+X^2_1+\cdots+X^2_1X^2_2\cdots X^2_k\Big)\,|u|^2\,dr + \int_{X_{k+1}^{-1}(R)}^{+\infty} |g'_{k+1}|^2\,dt\;.
$$

The asymptotical optimality shared at every order by the functions defined in~(\ref{BBB}) follows from the fact that for every $A>0$, 
$$
\inf_{g\in{\cal D}(A,+\infty)\,,\; g\not\equiv 0}\;\frac{\int_A^{+\infty} |g'|^2\,dt}{\int_A^{+\infty} |g|^2\,dt}\,= \, 0\;.
$$ 
Hence, there exists functions $u$ such that the first term in the r.h.s. of (\ref{oppt}) is negligible w.r.t. the second one. \finprf

\begin{corollary}\label{C1} Let $W:\R^+\to \R^+$ be a radially symmetric continuous and a.e. differentiable function satisfying (\ref{rpetit}). Then, the optimal asymptotic growth at the origin for all functions $W$ for which (\ref{R8}) holds in $H^1_0 (B_R,\C^2)$ is that of the function
\bq\label{BBBBB}
{W_\infty} (s)=1+\frac{1}8 \Big( \Sum_{j=1}^{+\infty} X_1^2(s)\cdots X_j^2(s) \Big)\;.
\eq 
\end{corollary}
\proof If $W$ violates the asymptotics given by (\ref{BBBBB}), a tedious calculation using~(\ref{variablechange}) shows that the corresponding potential $V$ violates the optimal asymptotics given by (\ref{BBB}). \finprf

\subsection{Optimal functions}\label{S3-3}

The first part of Theorem \ref{TR8} is proved by Lemma \ref{PP2}, Proposition \ref{Prop3} and Corollary \ref{C1}. For the second part, we have to match optimal functions near the origin and near infinity. 

\noindent 1) According to Corollaries \ref{CC1} and \ref{C1}, for $\bar n:=\frac{1}2 \Sum_{j=1}^{+\infty} X_1(s)\cdots X_j(s)$,
$$
\bar W:= \min(W^{+,1+\bar n}, W^{-,1-\bar n})= 1+\frac{1}{8}\Sum_{j=1}^{+\infty} X_1^2(s)\cdots X_j^2(s)+o(s)\quad\mbox{as}\; s\to 0^+
$$
is optimal near the origin (see Appendix A for more details). A simple computation shows that $\bar W$ becomes smaller than $1$ for any $s>R$, for some $R\in (0,1)$. A first example of a function $W\geq 1$ which has optimal behavior near the origin is therefore given by $W_1:=\max(\bar W,1)=\min( W^{+,\bar m_+}, W^{-, \bar m_-})$, with
$$
\bar m_\pm=\left\{\begin{array}{ll}1\pm\bar n&\quad\mbox{if}\; s<R_\pm\,, \\
1&\quad\mbox{if}\;s\geq R_\pm\,,\\ \end{array} \right.
$$
where $[0, R_\pm]$ is the support of $\bar W^{\pm, 1\pm \bar n}-1$.

\noindent 2) On the other hand, if we compute 
$$
\tilde W:= \min(W^{+,1+\tilde n}, W^{-,1-\tilde n})\quad\mbox{with}\quad\tilde n:= \frac 1{4\,s} -1\;,
$$
we notice that $\tilde W\geq 1$ for all $s\geq T$ for $T=\frac 1{48}[{\scriptscriptstyle{{\left( 4096 - 192\,{\sqrt{417}} \right) }^{1/3} + 4\,\left( 4 + {\left( 64 + 3\,{\sqrt{417}} \right) }^{1/3} \right)}}]$. Numerically, one finds $T\approx 0.866876$\dots\ Hence, $W_2:=\max(1, \tilde W)\geq 1$ is an example of a function $W\geq 1$ which has an optimal behavior at infinity : $W(s)\sim s$ as $s\to +\infty$. Note that $W_2=\min(W^{+, \tilde m_+}, W^{-, \tilde m_-})$, with 
$$
\tilde m_\pm=\left\{\begin{array}{ll}1&\quad\mbox{if}\; s<T_\pm\;,\\
1\pm\tilde n&\quad\mbox{if}\; s\geq T_\pm\;,\\ \end{array} \right.
$$
where $[T_\pm, +\infty)$ is the support of $ W^{\pm, 1\pm\tilde n}-1$.

\smallskip The function $W_2$ has an additional nice property : since for $s$ large, $W_2\approx s+\frac 1{8\,s}$, if we scale Inequality (\ref{R8}) keeping the $L^2$-norm constant, on one end of the scale we obtain Inequality \eqn{R2}, while on the other end we find the uncertainty principle / classical Hardy inequality (\ref{uncertainty}).

\noindent 3) Now we prove that one can optimize the behavior of $W$ near $0$ and near infinity simultaneously, with $W>1$ on $(0,+\infty)$.

\noindent\underline{Case $+$ :} We take $a$ large enough so that the function $W^+\kern -4pt:=\max({W_\infty}, W^{+, 1+\tilde n})$ is well defined, continuous in $(0,+\infty)$ and satisfies :
$$\begin{array}{ll}
W^+\equiv W_\infty&\quad\mbox{in}\; [0,R]\;,\\
W^+\equiv W^{+, 1+\tilde n}&\quad\mbox{in}\; [R,+\infty)\;,\\
W^+(R)>1\;,
\end{array}$$
for some $R>0$ (numerically, $a>5$ is enough). This amounts to define $W^+$ as $W^{+,m_+}$, with
$$ 
m_+ (s) \ = \ \left \lbrace\begin{array}{ll}
m&\quad\mbox{if}\; s \leq R\;,\\ 
W^{+, 1+\tilde n}&\quad\mbox{if}\; s\geq R\;,\\
\end{array} \right. 
$$
where $m$ is the solution of the O.D.E. problem
$$
s\,m' = -2m+s\,(m^2-1)+{ W_\infty}(1+m^2)\;,\quad m(0)=1\;.
$$
The existence of $m$ is proved in Appendix B.

\noindent\underline{Case $-$ :} This is dealt with in the same manner, by patching this time $W_\infty$ and $W^{-, 1-\tilde n}$ in an appropriate way.

\smallskip The function $W_3:=\min(W^-, W^+)$ satisfies all the properties stated in Theorem \ref{TR8}.

\medskip In all the above examples (where $W\not\equiv 1$) the functions $m_\pm$ have discontinuities and $C(R)<0$. Indeed, this has to be the case whenever $W\not\equiv 1$. Let $m_\pm$ be defined by (\ref{Wm}). According to Proposition \ref{PP3}, $m_\pm\geq 1$ in a neighbourhood of $s=0^+$. Using $W\geq 1$, we get
$$s\,m'_\pm\geq (m-1)^2\;,$$
and an easy O.D.E. argument shows that $m_\pm$ cannot be globally defined, so it must have a discontinuity. The arguments used in the proof of Proposition~\ref{PP1} allow us to conclude.\finprf

\section*{Appendix A : Properties of the functions $X_k$}

Let $a>1$. Define $X_1(s):=\frac 1{a-\log s}$ for any $s\in (0,e^{a-1})$, and, by induction for any $k\geq 1$, $X_{k+1}(s):=X_1(X_k(s))$. Note that
\[0<s<e^{a-1}\quad\Longrightarrow\quad 0<X_1(s)<1<e^{a-1}\;,\]
which implies that $s^*(a)=\lim_{k\to+\infty}X_k(s)\in (0,1)$ is independent of $s$ (the limit is unique since $d^2X_1/ds^2$ changes sign only once on $(0, e^a)$).
Then 
\[s\,\frac{dX_1}{ds}=X_1^2(s)\quad\mbox{and}\quad s\,X_{k+1}^{-1}\frac{dX_{k+1}}{ds}=X_{k+1}(s)\cdot s\,X_k^{-1}\frac{dX_k}{ds}\;.\]
Let $\displaystyle\pi_k(s):=\prod_{j=1}^kX_j(s)$ and $\displaystyle\sigma_k(s):=\sum_{j=1}^k\pi_j(s)$. Since $s\,X_{k+1}^{-1}\frac{dX_{k+1}}{ds}=\pi_{k+1}$, it follows that $s\,\frac{d\pi_k}{ds}=\pi_k\,\sigma_k$. By definition of $X_k$, $X_{k+1}(s)=X_k(t)_{|t=X_1(s)}$ and
\[\sigma_{k+1}(s)=t\Big(1+\sigma_k(t)\Big)_{|t=X_1(s)}\;\mbox{and}\; s\,\frac{d\sigma_{k+1}}{ds}=\left.\Big(t\,\frac{d\sigma_k}{dt}(t)+\sigma_k(t)+1\Big)t^2\right._{|t=X_1(s)}\,.\]
Using the two above identities, we can prove by induction the following formula~:
\begin{lemma} For any $k\geq 1$, for any $s\in (0,e^{a-1})$,
\[2s\,\frac{d\sigma_k}{ds}(s)-\sigma_k^2(s)=\sum_{j=1}^k\pi_j^2(s)\;.\]
\end{lemma}
We may now pass to the limit $k\to +\infty$. Let $\sigma(s):=\sum_{j=1}^{+\infty}\pi_j(s)$ :
\[2s\,\frac{d\sigma}{ds}(s)-\sigma^2(s)=\sum_{j=1}^{+\infty}\pi_j^2(s)\;.\]
With the notations $W_\infty:=1+\frac 18\,\sum_{j=1}^{+\infty}\pi_j^2(s)$ and $\bar n(s):=\frac 12\,\sigma=\frac 12\,\sum_{j=1}^{+\infty}\pi_j(s)$,
this means
\begin{corollary} For any $s\in (0,e^{a-1})$,
\[s\,\frac{d\bar n}{ds}-{\bar n}^2=2\,\Big(W_\infty-1\Big)\;.\]
\end{corollary}

\section*{Appendix B : Solving a singular O.D.E.}

Here we solve the differential equation
\bq\label{singode} 
W_\infty -1=\Frac{s\,n'-n^2-2\,s\,n-s\,n^2}{2+2n+n^2}\,,\quad n(0)=1\;,
\eq
in an interval $[0,\delta]$, $\delta>0$, small, with $W_\infty(s)\,=\,1\,+\,\Frac{1}{8}\,\Sum^\infty_{k=1}\,X_1(s)^2\cdots X_k(s)^2$. 
\begin{proposition} There exists $\delta>0$ such that \eqn{singode} has a continuous solution in $[0, \delta)$.\end{proposition}
Note that this problem is a limiting one in the sense that there is no function~$W$ more singular than $W_\infty$ at the origin, for which the above problem can be solved with continuity at the origin.

\proof Let $C$, $\delta$ be two positive constants and define the set 
$$
X_{C,\delta}:=\{u\in C([0,\delta])\;:\;\limsup_{s\to 0} |u(s)\,\log(s) |\leq C\}\;.
$$
Let us write $ n:= \bar n\,(1+w)=\frac{1}2 \sum_{j=1}^{+\infty} X_1(s)\cdots X_j(s)$. Then $n$ is a solution to~(\ref{singode}) if and only if $w$ is a solution to 
\bq\label{odew} 
w'= f_0+f_1\,w+f_2\,w^2\;,
\eq
where $f_0$, $f_1$, $f_2$ have the following behavior near $0$ :
\begin{eqnarray*}
&&f_0=\quad\mbox{\raisebox{-4pt}{$\widetilde{\scriptstyle s\to 0}$}}\quad\frac 1{4\,s\,|\log s|^2}\\
&&f_1=\quad\mbox{\raisebox{-4pt}{$\widetilde{\scriptstyle s\to 0}$}}\quad\frac 1{s\,|\log s|\,\log(|\log s|)}\\
&&f_2\quad\mbox{\raisebox{-4pt}{$\widetilde{\scriptstyle s\to 0}$}}\quad\frac 1{s\,|\log s|}\\
\end{eqnarray*}
In order to solve equation (\ref{odew}) together with the initial value $w(0)=0$, we introduce the map $T: X_{C,\delta}\to X_{C, \delta}$ defined by
$$
Tw(s):=\int_0^s (f_0+f_1\,w+f_2\,w^2)\,dy\;,
$$
and look for a fiwed point. By choosing $C>1/4$ and $\delta<1$ small enough, $T$ maps $X_{C,\delta}$ into itself and it is a contraction. So, there is a unique solution of~(\ref{odew}) in $X_{C,\delta}$ which means that \eqn{singode} has a unique continuous solution $n$ in the interval $[0,\delta]$), with $n/\bar n-1$ in $X_{C,\delta}$, such that $n(0)=0$.

\section*{Appendix C : Why do we have a $\liminf$ in Theorem \ref{TR8} ?}

We are going to give a qualitative example showing that only a $\liminf$ can be achieved. Let $W:=W_\infty+\Sum_{n\geq 0}\bar W\Big(\frac{s-s_n}{\varepsilon_n}\Big)$ where $\varepsilon_n$ and $s_n$ are such that 
$$\begin{array}{c}\displaystyle s_n>0\,,\quad\lim_{n\to+\infty}s_n=0\;,\\
\displaystyle\varepsilon_n>0\,,\quad\sum_{n\geq 0}\varepsilon_n<+\infty\;,\\
\displaystyle\varepsilon_{n+1}<s_n-s_{n+1}\;,\end{array}$$
and assume that $\bar W$ is a bounded function with compact support in $(0,1)$. Then 
$$\limsup_{s\to 0^+}W(s)>1$$
and the equations 
$$2m_\pm\pm s\,m_\pm'-s\,m_\pm^2+s=W(1+m_\pm^2)$$
have no solution continuous up to $s=0$.


\medskip\noindent{\bf Acknowledgment :} {\sl M.J.E. would like to thank the Georgia Tech School of Mathematics for its hospitality and M.L. would like to thank CEREMADE where some of this work has been carried out.}


\end{document}